\documentclass[apj,usenatbib,twocolumn]{aastex6}

\usepackage{float}
 \usepackage{color}
 \usepackage{xspace}
\usepackage{threeparttable}
\usepackage{nicefrac}
\usepackage{afterpage}
 \usepackage{url}
 \usepackage{hyperref}
 \usepackage{natbib}
\usepackage{todonotes}

\newcommand{\angstrom}{\textup{\AA}\xspace}
\def\kms{{\rm km\,s$^{-1}$}\xspace}
\def\ergs{{\rm\,erg\,s$^{-1}$}\xspace}
\def\Lsun{{\rm L$_{\odot}$}\xspace}
\def\Rsun{{\rm R$_{\odot}$}\xspace}
\def\Msun{{\rm M$_{\odot}$}\xspace}

\def\halpha{{\rm H$\alpha$}\xspace}
\def\he2{{He~{\small II}}\xspace}

\newcommand{\io}[2]{#1~{\small #2}}

\begin{document} 

   \title{\lowercase{i}PTF16\lowercase{fnl}: a faint and fast tidal disruption event in an E+A galaxy}

   \author{N. Blagorodnova \altaffilmark{1},
   			S.~Gezari \altaffilmark{2,3},
            T.~Hung \altaffilmark{2},
			S.~R.~Kulkarni \altaffilmark{1},
            S.~B.~Cenko \altaffilmark{3,4},
            D.~R.~Pasham \altaffilmark{5,26}$^{\dagger}$,
            L.~Yan \altaffilmark{1,6},
            I.~Arcavi \altaffilmark{7,8,26}$^{\dagger}$,
            S.~Ben-Ami \altaffilmark{9,10},
            B.~D.~Bue \altaffilmark{11},
            T.~Cantwell \altaffilmark{12},
            Y.~Cao \altaffilmark{13},
            A.~J.~Castro-Tirado \altaffilmark{14,15},
            R.~Fender \altaffilmark{16},
 			C.~Fremling \altaffilmark{17},
            A.~Gal-Yam \altaffilmark{10},
			A.~Y.~Q.~Ho \altaffilmark{1},
            A.~Horesh \altaffilmark{18},
			G.~Hosseinzadeh \altaffilmark{7,8},
            M.~M.~Kasliwal \altaffilmark{1},
            A.~K.~H.~Kong \altaffilmark{16,19},
            R.~R.~Laher \altaffilmark{6},
            G.~Leloudas \altaffilmark{10,20},
            R.~Lunnan \altaffilmark{1,17},
             F.~J.~Masci \altaffilmark{6},
            K.~Mooley \altaffilmark{16},
            J.~D.~Neill \altaffilmark{1},
             P.~Nugent \altaffilmark{21,22},
             M.~Powell \altaffilmark{23},
             A. F. Valeev \altaffilmark{24},
             P.~M.~Vreeswijk \altaffilmark{10},
             R.~Walters \altaffilmark{1},
             P.~Wozniak \altaffilmark{25}
            }

   \altaffiltext{1}{
   Cahill Center for Astrophysics, California Institute of Technology, Pasadena, CA 91125, USA; nblago@caltech.edu }
     \altaffiltext{2}{ Department of Astronomy, University of Maryland, Stadium Drive, College Park, MD 20742-2421, USA }
     \altaffiltext{3}{Joint Space-Science Institute, University of Maryland, College Park, MD 20742, USA }
     \altaffiltext{4}{NASA Goddard Space Flight Center, Mail Code 661, Greenbelt, MD 20771, USA}
     \altaffiltext{5}{ Center for Space Research, Massachusetts Institute of Technology, Cambridge, MA 02139}
          \altaffiltext{6}{Infrared Processing and Analysis Center, California Institute of Technology, Pasadena, CA 91125, USA}    
          \altaffiltext{7}{Department of Physics, University of California, Santa Barbara, CA 93106-9530, USA}
\altaffiltext{8}{Las Cumbres Observatory, 6740 Cortona Dr Ste 102, Goleta, CA 93117-5575, USA}

\altaffiltext{9}{Smithsonian Astrophysical Observatory, Harvard-Smithsonian Center for Astrophysics, 60 Garden St., Cambridge, MA 02138, USA}
     \altaffiltext{10}{ Department of Particle Physics and Astrophysics, Weizmann Institute of Science, Rehovot 7610001, Israel }
      \altaffiltext{11}{ Jet Propulsion Laboratory, California Institute of Technology, Pasadena, CA 91125, USA  }  
      \altaffiltext{12}{ Jodrell Bank Centre for Astrophysics, Alan Turing Building, Oxford Road, Manchester M13 9PL, UK } 
      \altaffiltext{13}{Department of Astronomy, University of Washington, Box 351580, Seattle, WA 98195-1580, USA}
	\altaffiltext{14}{Instituto de Astrof\'isica de Andaluc\'ia (IAA-CSIC), P.O. Box 03004, E-18080 Granada, Spain}
	\altaffiltext{15}{Unidad Asociada Departamento de Ingeniería de Sistemas y Automática, Univ. de M\'alaga, Spain}
	\altaffiltext{16}{ Astrophysics, Department of Physics, University of Oxford, Keble Road, Oxford OX1 3RH, UK}  
	\altaffiltext{17}{ Department of Astronomy, The Oskar Klein Center, Stockholm University, AlbaNova, 10691 Stockholm, Sweden }     
    \altaffiltext{18}{Racah Institute of Physics, Hebrew University, Jerusalem, 91904, Israel}
    \altaffiltext{19}{Institute of Astronomy and Department of Physics, National Tsing Hua University, Hsinchu 30013, Taiwan}
        \altaffiltext{20}{ Dark Cosmology centre, Niels Bohr Institute, University of Copenhagen, Juliane Maries vej 30,  2100 Copenhagen, Denmark }
	\altaffiltext{21}{ Lawrence Berkeley National Laboratory, Berkeley, California 94720, USA }
        \altaffiltext{22}{Department of Astronomy, University of California, Berkeley, CA 94720-3411, USA}
\altaffiltext{23}{Department of Physics and Yale Center for Astronomy \& Astrophysics, Yale University PO Box 208120, New Haven, CT 06520-8120, USA}
     \altaffiltext{24}{Special Astrophysical Observatory, Nizhnij Arkhyz, Karachai-Cherkessian Republic, 369167 Russia}
    \altaffiltext{25}{ Los Alamos National Laboratory, MS-D466, Los Alamos, NM 87545, USA }
       \altaffiltext{26}{$^\dagger$Einstein Fellow}

\date{Received XX XX XXXX; accepted XX XX, XXXX}

  \begin{abstract}
  
  We present ground-based and \textit{Swift} observations of iPTF16fnl, a likely tidal disruption event (TDE) discovered by the intermediate Palomar Transient Factory (iPTF) survey at 66.6\,Mpc. The lightcurve of the object peaked at absolute $M_g=-17.2$\,mag. The maximum bolometric luminosity (from optical and UV) was $L_p~\simeq~(1.0\,\pm\,0.15) \times 10^{43}$\ergs, an order of magnitude fainter than any other optical TDE discovered so far. The luminosity in the first 60 days is consistent with an exponential decay, with $L \propto e^{-(t-t_0)/\tau}$, where $t_0$=~57631.0 (MJD) and $\tau\simeq 15$\,days. The X-ray shows a marginal detection at $L_X=2.4^{1.9}_{-1.1}\times 10^{39}$\,\ergs (\textit{Swift} X-ray Telescope). No radio counterpart was detected down to 3$\sigma$, providing upper limits for monochromatic radio luminosity of $\nu L_{\nu} < 2.3\times10^{36}$ \ergs and $\nu L_{\nu}<1.7\times 10^{37}$\,\ergs (VLA, 6.1 and 22\,GHz). The blackbody temperature, obtained from combined \textit{Swift} UV and optical photometry, shows a constant value of 19,000\,K. The transient spectrum at peak is characterized by broad \he2 and \halpha emission lines, with an FWHM of about 14,000\,\kms and 10,000\,\kms respectively. \io{He}{I} lines are also detected at $\lambda\lambda$ 5875 and 6678. The spectrum of the host is dominated by strong Balmer absorption lines, which are consistent with a post-starburst (E+A) galaxy with an age of $\sim$650\,\,Myr and solar metallicity. The characteristics of iPTF16fnl make it an outlier on both luminosity and decay timescales, as compared to other optically selected TDEs. The discovery of such a faint optical event suggests a higher rate of tidal disruptions, as low luminosity events may have gone unnoticed in previous searches.
\end{abstract}

   \keywords{accretion, accretion discs  -- black hole physics -- stars: individual (iPTF16fnl), galaxies: nuclei  }

\maketitle

\section{Introduction}

A tidal disruption event (TDE) is the phenomenon observed when a star is torn apart by the tidal forces of a supermassive black hole (SMBH), usually lurking in the core of its galaxy.  As a consequence, a bright flare is expected when some of the bound material accretes onto the SMBH. Although such events were theoretically predicted a few decades ago \citep{Hills1975Natur,Lacy1982,Rees1988,EvansKochanek1989ApJ,Phinney1989IAUS}, observational signatures are more recent. The first detection of TDEs were made in the soft X-ray data. The flares, consistent with the proposed stellar disruption scenario, were identified in ROentgen SATellite (ROSAT) all-sky survey \citep{Bade1996AA,Komossa1999AA,Saxton2012AA}. Detection in gamma-ray data of \textit{Swift} events 
Swift J1644+75 \citep{Bloom2011Sci,Levan2011Sci,Burrows2011Natur,Zauderer2011Natur,Cenko2012ApJ} and Swift J1112.2-8238 \citep{Brown2015MNRAS} were attributed to relativistic outbursts caused by jetted emission. We refer the reader to \cite{Komossa2015,Auchettl2016arXiv} for a broader review of the status of observations in different wavelengths. 

\begin{figure*}[!ht]
\includegraphics[width=\textwidth]{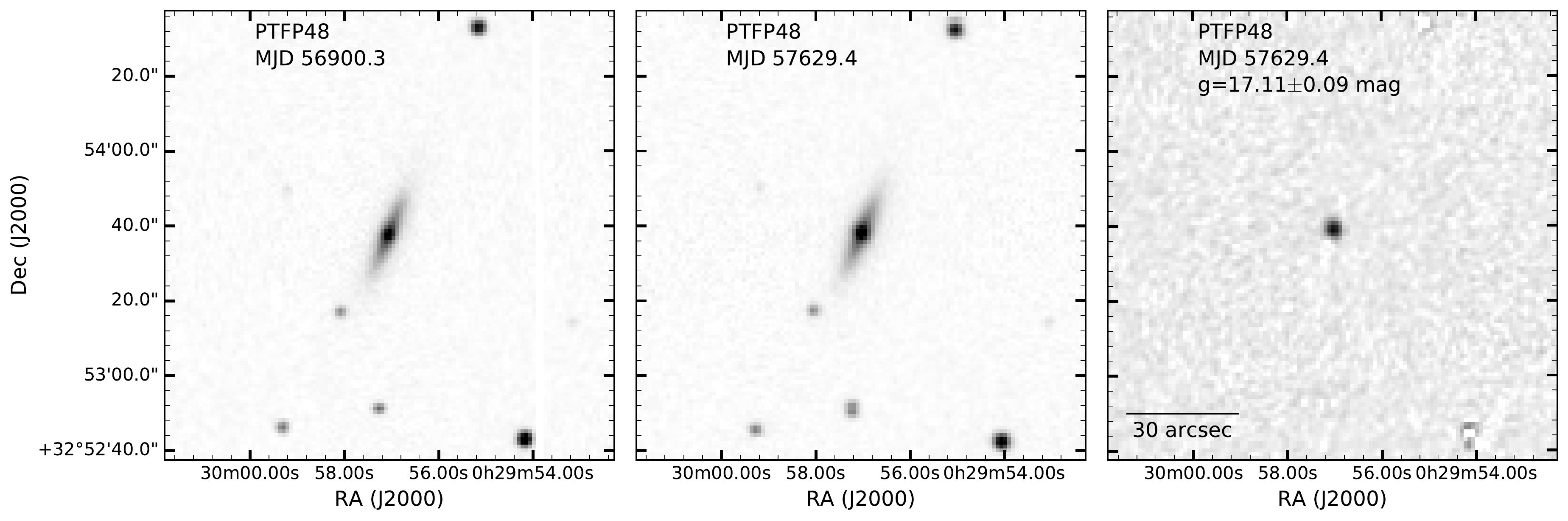} 
\caption{P48 cutouts of 2$'$ diameter sky region centered in the position of the transient. The coordinates of the transient are
$\alpha_{J2000}=00^{\rm{h}}29^{\rm{m}}57^{\rm{s}}.05$,  $\delta_{J2000}=+32^{\rm{h}}53^{\rm{m}}37^{\rm{s}}.48$. The cutouts show the host galaxy approximately one year before the discovery, on the day of the discovery, and the difference image showing the transient location in the core of the galaxy. }
\label{fig:finder}
\end{figure*}

Ultraviolet detections of nuclear flares were reported from the \textit{GALEX} survey \citep{Gezari2006ApJ,Gezari2008ApJ,Gezari2009ApJ}. TDEs are now being discovered by optical surveys such as the Sloan Digital Sky Survey \citep[SDSS;][]{vanVelzen2011ApJ}, PanSTARRS-1 \citep[PS1;][]{Gezari2012Natur,Chornock2014ApJ}, Palomar Transient Factory \citep[PTF;][]{Arcavi2014}, All Sky Automated Survey for Supernovae \citep[ASAS-SN;][]{Holoien2014-14ae,Holoien2016-14li,Holoien2016-15oi}, Robotic Optical Transient Search Experiment \citep[ROTSE][]{Vinko2015ApJ}, Optical Gravitational Lensing Experiment \citep[OGLE;][]{Wyrzykowski2017MNRAS} and the intermediate Palomar Transient Factory (iPTF; \cite{Hung2017arXiv}, Duggan G. et. al. in prep and the work presented here). The optical sample has revealed that an important fraction of the TDEs appear to be found in E+A (``quiescent Balmer-strong'') galaxies \citep{Arcavi2014,French2016ApJ}, which can be interpreted as middle-aged ($<1$\,Gyr) post-starburst galaxies \citep{Zabludoff1996ApJ,Dressler1999ApJS}.

Here we present  iPTF16fnl, an optical TDE candidate discovered by iPTF. The event is localized in the center of an E+A galaxy. With a distance of 66.6\,Mpc this is the closest well-studied event in optical/UV wavelengths. 

\section{Discovery and Host Galaxy}\label{sec:DiscoveryHostGalaxy}
\subsection{Discovery and classification}

iPTF16fnl was discovered on UT 2016 August 29.4 in an image obtained during the $g$+R experiment \citep{Miller2017arXiv}: during the night, one image each was obtained in $g$ and R band; the images are separated by at least an hour in order to filter out asteroids. The event was identified by two real-time difference imaging pipelines \citep{Cao2016PASP,Masci2017PASP}, shown in Figure \ref{fig:finder}. The discovery magnitudes were $g=17.11\pm0.09$ and $R=17.39\pm0.09$ (see \cite{Ofek2012PASP} for photometric calibration of PTF). Given an RMS of 0.5$''$, the coordinates of the source, $\alpha_{J2000}=00^{\rm{h}}29^{\rm{m}}57^{\rm{s}}.04\ \delta_{J2000}=+32^{\rm{h}}53^{\rm{m}}37^{\rm{s}}.5$ (ICRS) are consistent with central position of the galaxy, as provided by the Sloan Digital Sky Survey (SDSS) catalogue \citep{SDSSDR12}.

\begin{figure}
\includegraphics[width=0.5\textwidth]{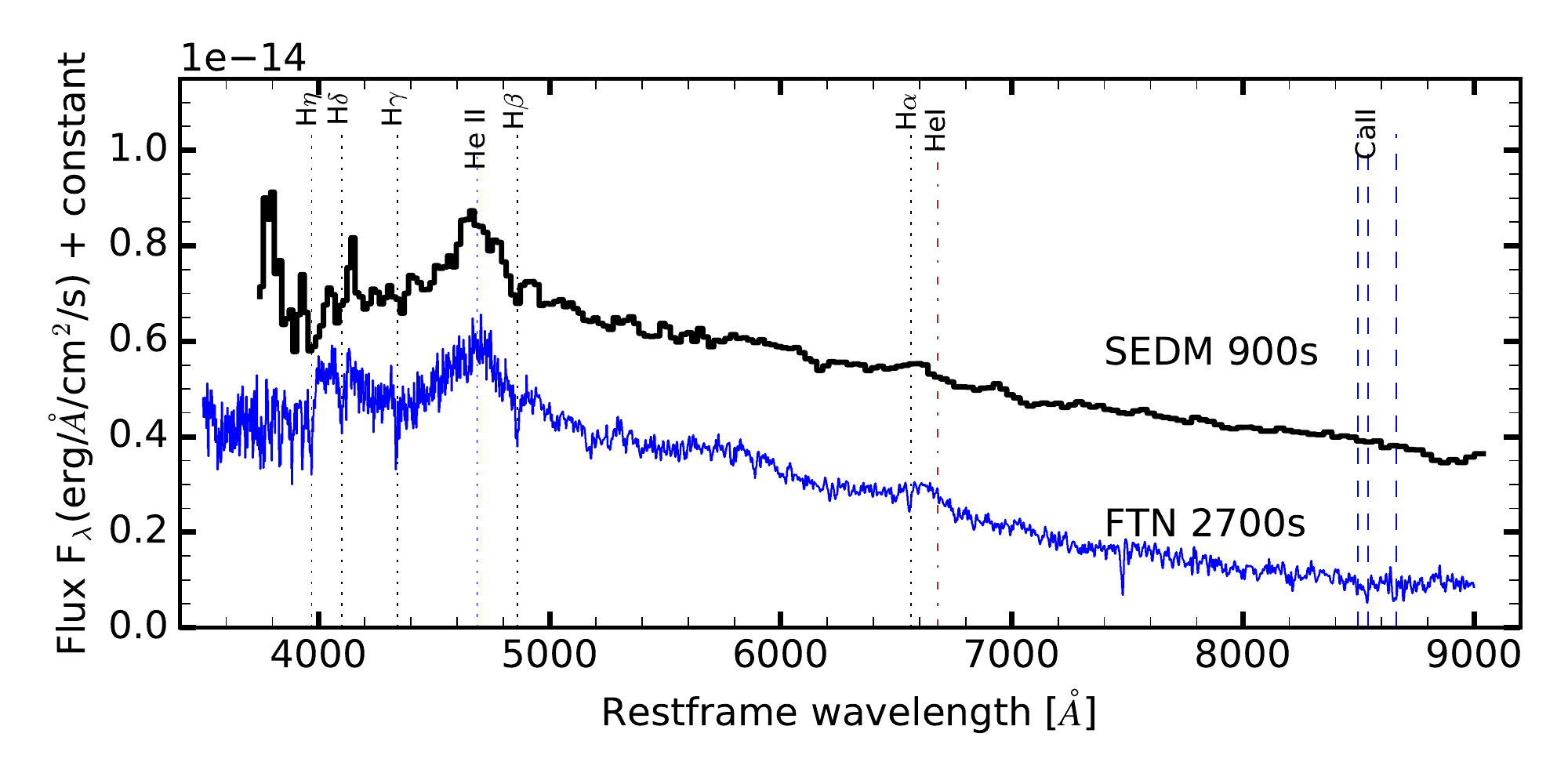} 
\caption{Two classification spectra obtained 2\,days before the peak in $g$-band. The top black thick line shows the 900\,s exposure obtained with SEDM, on the Palomar 60-inch (1.5\,m). The bottom blue line shows a 2700\,s exposure obtained with the FLOYDS spectrograph on the Las Cumbres Observatory (LCO) 2-m telescope. The blue continuum and prominent \he2 line are clearly identified in both spectra.}
\label{fig:discovery}
\end{figure}

The brightness, blue colour ($g-{\rm R}=-0.28$\,mag) and central location of the transient in its host galaxy, made it a prime candidate for prompt follow-up observation. On the night after discovery, we observed the source with the FLOYDS spectrograph \citep{FLOYDS2011AAS} on the Las Cumbres Observatory \citep[LCO;][]{Brown2013PASP} 2-m telescope  and the Spectral Energy Distribution Machine (SEDM) on the Palomar 60-inch (P60) telescope. FLOYDS are a pair of robotic low-resolution (R $\sim$ 450) slit spectrographs optimized for supernova classification and follow-up. The SEDM is a ultra-low resolution (R$\sim100$) integral-field-unit (IFU) spectrograph, dedicated to fast turnaround classification (Blagorodnova N. et. al. in prep). The IFU's wide field-of-view of 28$''$, allows for robotic spectroscopy. Their data reduction pipeline allow rapid reduction with minimal intervention from the user (aperture placing). Figure \ref{fig:discovery} shows the classification spectra, displaying a blue continuum and broad \he2 and H$\alpha$ emission lines, characteristic of previously observed optical TDE spectra \citep{Arcavi2014}.  The fast spectral identification of iPTF16fnl as a TDE candidate allowed us to rapidly inform the astronomical community  \citep[ATel \#9433;][]{Gezari2016ATel}, which in turn enabled numerous multi-wavelength follow-up campaigns.

\subsection{Host galaxy} \label{sec:host}

The host galaxy of iPTF16fnl is Markarian\,950 (Mrk 950) located at $z$=0.016328 \citep{Cabanela1998,Petrosian2007}. Given the edge-on host inclination and the existence of a peculiar bar and nucleus, the galaxy is classified as Sp (\textit{spindle}). The luminosity distance is D$_L$=66.6\,Mpc  (distance modulus $\mu=$~34.12\,mag) using H$_0$ =  69.6\,\kms\,Mpc$^{-1}$, $\Omega_M$ =   0.29, $\Omega_{\Lambda}$ = 0.71 in the reference frame of the 3\,K cosmic microwave background \citep[CMB;][]{CMB1996ApJ}.  

The estimated Galactic colour excess at the position of the transient is $E({\rm B}-{\rm V})~=~0.062~\pm~0.001$\,mag (from NED\footnote{The NASA/IPAC Extragalactic Database (NED) is operated by the Jet Propulsion Laboratory, California Institute of Technology, under contract with the National Aeronautics and Space Administration.}), after adopting the extinction law of \cite{Fitzpatrick1999PASP} with corrections from \cite{Schlafly2011ApJ}. Assuming R$_{\rm V}~=~3.1$, the Galactic visual extinction is A$_{\rm V}$ = 0.192\,mag.

\begin{table}
\begin{minipage}{1.\linewidth}
\begin{small}
\caption{Archival photometry of Mrk 950.}
\centering
\begin{tabular}{cccccccc}
\hline
Survey & Band  & Magnitude & Reference    \\ 
    &  		& (mag) &     \\ \hline
 GALEX 		& FUV$_{AB}$ & 21.22$\pm$0.39 $^a$& [1] \\
 GALEX 		& NUV$_{AB}$ & 20.191$\pm$ 0.13 $^a$& [1] \\
 SDSSDR12 	& $u$ & 17.005$\pm$0.012  $^b$& [2]\\
 SDSSDR12 	& $g$ & 15.491$\pm$0.003  $^b$& [2]\\
 SDSSDR12 	& $r$ & 14.913$\pm$0.003  $^b$& [2]\\
 SDSSDR12 	& $i$ & 14.585$\pm$0.003  $^b$& [2]\\
 2MASS 		& $J$ & 13.212	$\pm$0.040 & [3] \\
 2MASS 		& $H$ & 12.545$\pm$0.052 & [3] \\
 2MASS 		& $K$ & 12.360$\pm$0.069 & [3] \\
 WISE 		&	W1	&	13.075$\pm$0.029&	[4] \\
 WISE		&	W2	&13.086$\pm$0.034&	[4] 	\\
 WISE		&	W3	&12.331$\pm$0.285&	[4] 	\\
 WISE		&	W4	&$>$9.136	&		[4]  \\  \hline
 NVSS		& 1.4\,GHz & $<$1.7 mJy& [5]\\
 ROSAT		&0.1-2.4 keV& $<$1.04$\times 10^{-12}\,$\ergs		& [6]\\
 \hline
 
\hline
\end{tabular}
\begin{tablenotes}
    \item[\textdagger]$^a$ Measured within 7.5$^{\prime\prime}$ diameter aperture. $^b$ Model magnitude. References: [1]  \cite{Bianchi2011}, [2] \cite{SDSSDR12}, [3] \cite{Jarrett20002MASS}, [4] \cite{WISE2010}, [5] \cite{Condon1998AJ}, [6] \cite{Voges1999AA}
    \end{tablenotes}
\label{tab:hostmag}
\end{small}
\end{minipage}
\end{table}

The archival host magnitudes are shown in Table \ref{tab:hostmag}. The derived $K$-band luminosity of the galaxy is  $L_K=1.15 \times 10^{10}$\Lsun. Using the galaxy color $u-g=1.5$, we compute a mass to light ratio for the $K$-band, log$_{10}(M/L)=-0.0755$ \citep{Bell2003ApJS}, corresponding to a stellar mass $M_{*} \simeq 9.7 \times 10^{9}$\Msun. 
Provided that size of the PSF in SDSS DR13 is an upper limit for the angular size of an unresolved bulge, we assume that an upper limit for the bulge-to-total ratio (B/T) can be estimated with an average (across all bands) value of \texttt{psfFlux}/\texttt{cModelFlux}$ \sim 0.19 \pm0.05$. We use this ratio to scale the total galaxy mass and derive $M_{BH} \leq 10^{6.6 \pm (0.1 + 0.34)} $\,\Msun, according to the $M_{BH} -M_{\rm{bulge}}$ relation \citep{McConnell2013ApJ}, including the 1$\sigma$ scatter of 0.34 dex.

The age and metallicity of the host galaxy were determined by fitting a grid of galaxy models from \cite{BruzualCharlot2003MNRAS}, using stellar evolutionary models from \citep{Chabrier2003PASP}. The fit was done using the \texttt{pPXF} code \citep{2016arXiv160708538C}. The best model was in agreement with a single burst of star formation with an age of 650$\pm$300\,Myr and a metallicity of $Z$=0.18 (here, $Z=0.2$ corresponds to that of the Sun).

We use the high resolution spectrum taken with VLT/UVES taken two weeks after discovery (see log in Table \ref{tab:speclog}, PI: P. Vreeswijk) to measure the host velocity dispersion. We obtain $v_{\rm{host}}=89\pm $1\,\kms, from the \io{Ca}{II} $\lambda\lambda$8544,8664 absorption lines. According to the M$-\sigma$ relation \citep{McConnell2013ApJ}, this corresponds to a $M_{BH}=10^{6.33 \pm 0.38}$\,\Msun, consistent with our previous estimate.

The field of iPTF16fnl was extensively observed for the last six years by PTF/iPTF. No prior activity in the host is detected with upper limits of $\sim$20-21 mag. The most recent non-detection is from MJD 57432.6, around 194 days before discovery. The host galaxy was also monitored by The Catalina Real-Time Transient Survey \citep[CRTS;][]{Drake2009ApJ} during  2006-2016 period. The magnitude of the galaxy was stable within the errors, with an unfiltered average magnitude of 14.88$\,\pm\,$0.05 mag.

\section{Follow-up observations} \label{sec:followup}

\subsection{Photometric observations}\label{sec:photometry}

Following spectroscopic identification of iPTF16fnl as a TDE candidate, the source was monitored at Palomar and the Ultraviolet and Optical Telescope \citep[UVOT;][]{Roming2005}  on board the \textit{Swift} observatory \citep{Gehrels2004ApJ}. The UVOT  observations were taken in \textit{UVW2}, \textit{UVM2}, \textit{UVW1}, \textit{U}, \textit{B} and \textit{V}; see Table~\ref{table:phot}. The data were reduced using the software \texttt{UVOTSOURCE} using the calibrations described in \cite{Poole2008} and updated calibrations from \cite{Breeveld2010MNRAS}. We use a 7.5$''$ aperture centered on the position of the transient. 

At Palomar, photometry in the $g$ and Mould-$R$ bands were obtained with the iPTF mosaic wide-field camera on the Palomar 48-inch telescope \citep[P48;][]{Rahmer2008}. Difference-image photometric measurements were provided by the IPAC Image Subtraction and Discovery Pipeline developed for the iPTF survey \citep[PTFIDE;][]{Masci2017PASP}. Difference-imaging photometry in the \textit{u'g'r'i'} bands, obtained with the SEDM, were computed using the \texttt{FPipe} software \citep{Fremling2016}. The zeropoints were calibrated using stars in the SDSS footprint. 
Table \ref{table:phot} reports the measured \textit{Swift} aperture photometry magnitudes and the difference imaging photometry for the Palomar data. 

\begin{figure*}[ht]
\includegraphics[width=2.1\columnwidth]{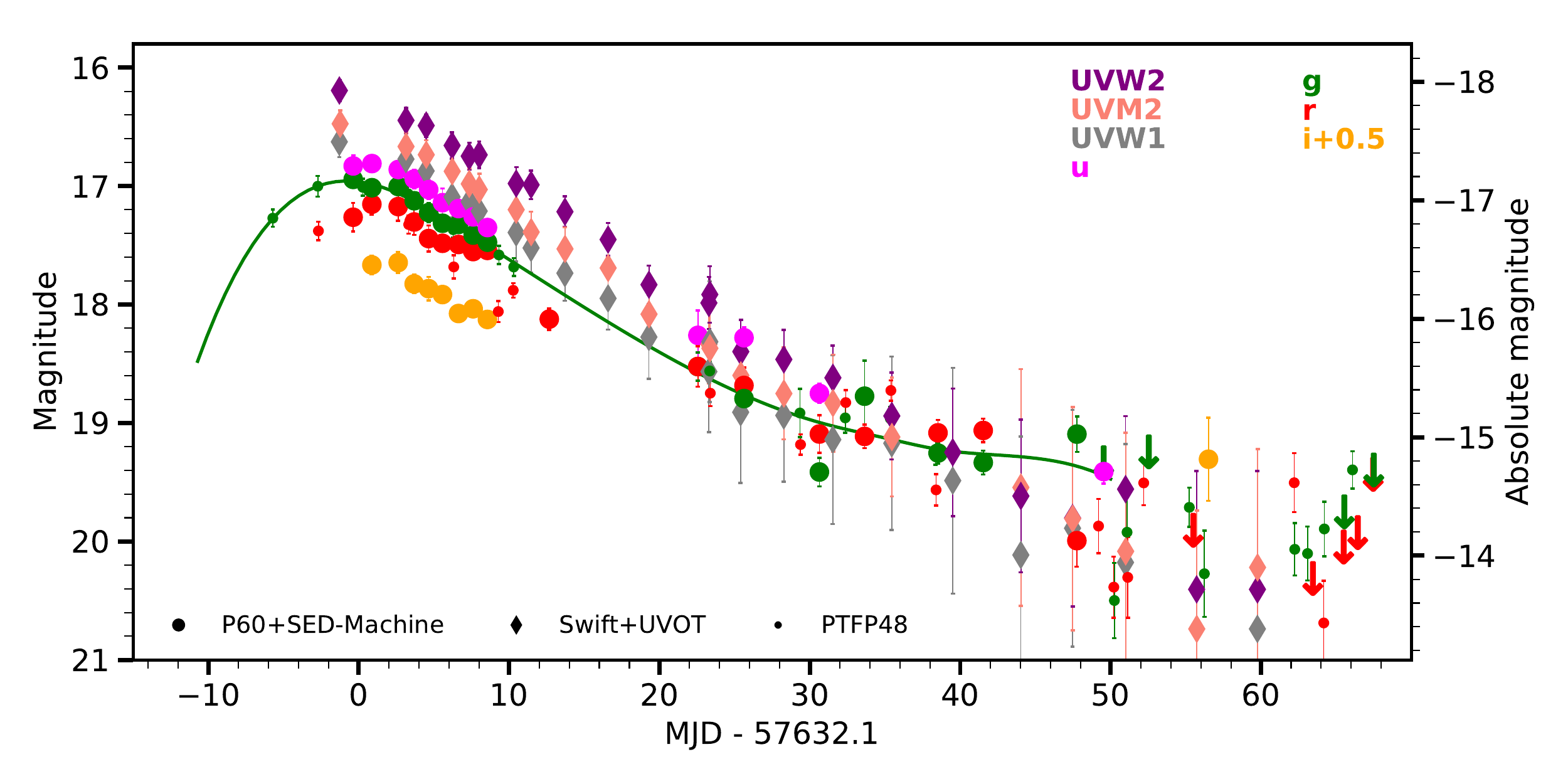} 
\caption{ Observed lightcurve for iPTF16fnl. The green solid line shows the best fit spline to the $g$-band data, which was corrected for Galactic extinction. The errors for epochs later than 30 days are likely underestimated, as the bulge of the host is $\sim$4 magnitudes brighter than the transient. The time of peak in $g$-band, MJD 57632.1 is used as the reference epoch.
The small symbol ``S'' on top shows the epochs when spectra were taken. (A color version of this figure is available in the online journal.)
}
\label{fig:lc}
\end{figure*}

The multi-band \textit{Swift} and optical lightcurve, corrected for Galactic extinction, is shown in Figure \ref{fig:lc}. To correct the UV bands for host light contamination, we used the average of the last four epochs of \textit{Swift} data, from MJD 57712.7 to 57724.7 ($>$ 80\,days), to subtract from the early part of the lightcurve. The measurements were constant with an RMS of $\sim$0.05 mag, comparable to the measurement error. Optical \textit{Swift} data were excluded from the analysis, as they were dominated by the host, not by the central point source.

In order to estimate the extinction in the host galaxy, we use all available \textit{Swift} UV data and difference imaging photometry in $ugri$ bands to fit blackbody emission curves, as detailed in Section \ref{sec:analysis}. For each epoch, the photometry is corrected for both Galactic (fixed) and additional host extinction E($B-V$) from 0 to 0.25\,mag in 0.05 steps. The likelihood between the de-reddened photometry and a blackbody model is computed for each epoch. In a final step, we marginalize over all epochs to derive the final value. We find that the best fit corresponds to E($B-V$)=0, with an upper limit of E($B-V$)=0.05. The selection of different extinction laws for host extinction \citep{Calzetti2000ApJ,Fitzpatrick1999PASP} does not change our conclusions. Given the assumption that the emission from iPTF16fnl in fact follows a  distribution, from now on we will assume the reddening in the host to be negligible. 


\subsection{Radio observations}

\begin{figure}[hb!]
\includegraphics[width=0.5\textwidth]{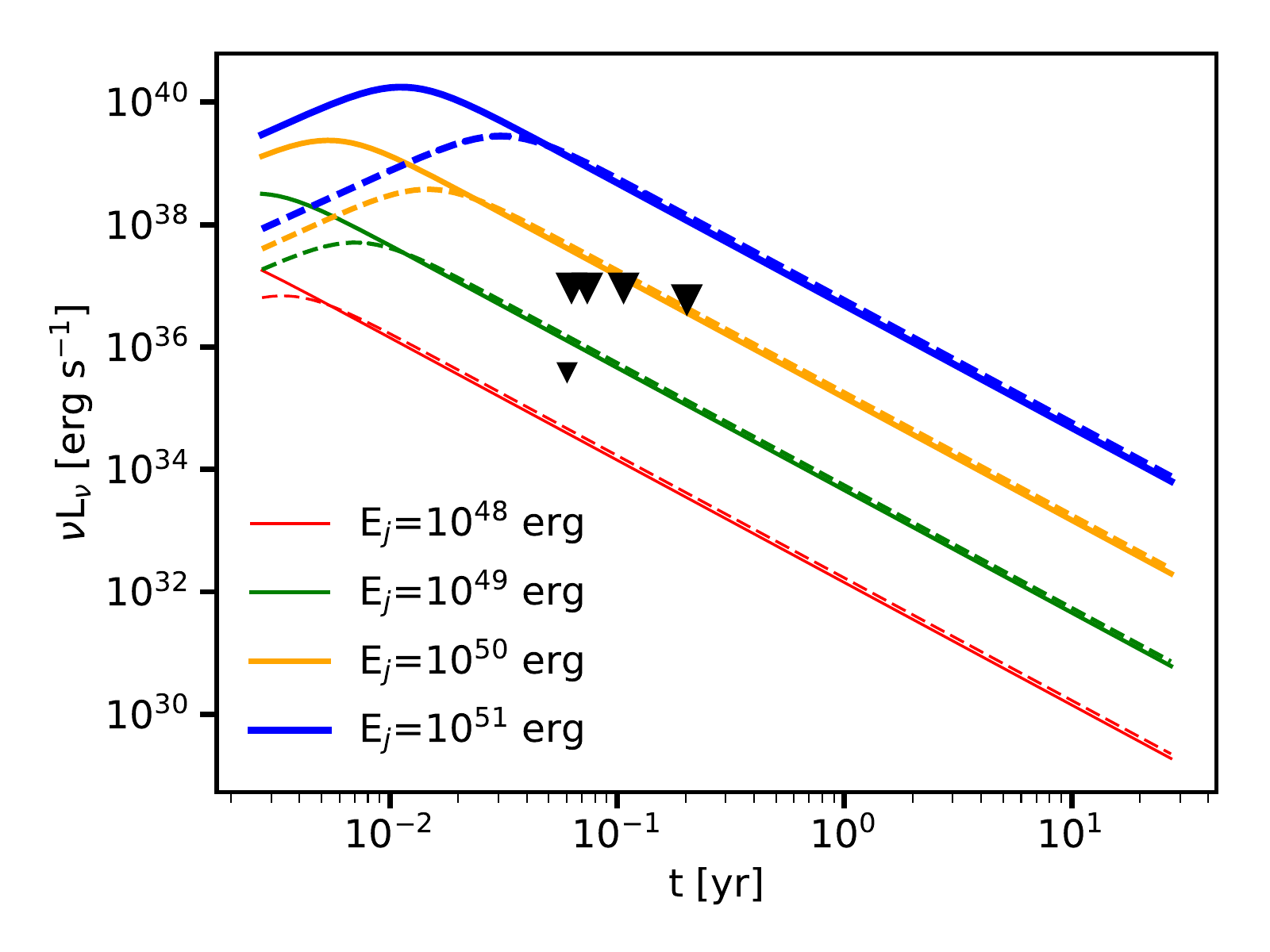} 
\caption{Upper limits for our 6\,GHz (small triangle) and 15\,GHz (big triangles) observations. The lines show analytic lightcurves for different TDE on-axis jet energies (colour coded) from \cite{Generozov2017MNRAS}. Solid lines represent the lightcurves for 15\,GHz and dashed lines for 6\,GHz. We assume $n_{18}=11$ and the fiducial values provided in their models for an optically thick case. The results are also consistent with an optically thin case. The $X$-axis is computed relative to our lower limit of 11 days for the time to peak light. }
\label{fig:radio}
\end{figure}

\begin{table}
\begin{minipage}{1.\linewidth}
\begin{small}
\caption{3$\sigma$ upper limits on radio emission for iPTF16fnl.}
\centering
\begin{tabular}{cccccccc}
\hline
Date 			& Telescope & Frequency		& Flux \\ 
 (UT)   		& 			&  		(GHz) 	& (Jy) \\ \hline
2016 Aug 31 	& VLA 		& 6.1, 22 		& $<12\times 10^{-6}$ \\
2016 Sep 01 &	AMI		&	15& $<117\times 10^{-6}$ \\
2016 Sep 05 &	AMI		&	15& $<117\times 10^{-6}$ \\ 
2016 Sep 09 &JCMT/SCUBA2&	352 		& $<10\times 10^{-3}$ \\
2016 Sep 10 &JCMT/SCUBA2&	352 		& $<7.5\times 10^{-3}$ \\
2016 Sep 17 &	AMI		&	15& $<117\times 10^{-6}$ \\
2016 Oct 22 & AMI		&	15& $<75\times 10^{-6}$	\\

\hline
\end{tabular}
\label{tab:radio}
\end{small}
\end{minipage}
\end{table}

Radio follow-up observations of iPTF16fnl were taken with the Jansky Very Large Array (VLA; PI A. Horesh), the Arcminute Microkelvin Imager (AMI; PI K. Mooley) and the James Clerk Maxwell Telescope and the Submillimetre Common-User Bolometer Array 2 (JCMT/SCUBA-2; PI A. K. H. Kong). The upper limits corresponding to the observations are shown in Table \ref{tab:radio}. The limits for the monochromatic radio luminosities corresponding to the first VLA epoch are $\nu L_{\nu} < 2.3\times10^{36}$ \ergs and $\nu L_{\nu} < 1.7\times 10^{37}$\,\ergs at 6.1 and 22\,GHz. These limits are respectively two and one order of magnitude deeper than the VLA detection of ASASSN-14li at peak for a similar frequency range \citep{vanVelzen2016Sci,Alexander2016ApJ}. 

We can use the limits reported here to argue against the presence of an on-axis relativistic outflow, or at least constrain the energy of the jet, $E_j$. We compare our limits in 6 and 15\,GHz with the analytical lightcurves for on-axis TDE radio emission from  \cite{Generozov2017MNRAS}. Figure \ref{fig:radio} shows that our early time observations suggest jet energies with $E_j< 10^{49}$\,erg\,s$^{-1}$.

Additionally, we contrast our measurements, scaled to the redshift of PS1-11af, with the GRB afterglow models of \cite{vanEerten2012ApJ} presented in \cite{Chornock2014ApJ} (their figure 13). Based on our non-detections, we can rule out the existence of a relativistic jet, as viewed 30$^{\circ}$ off-axis. For larger angles, the radio emission is expected to arise at later times ($>$ 1 year after disruption). Therefore, continuous monitoring of the event at radio wavelengths is encouraged.

\subsection{X-ray observations}

We observed the location of iPTF16fnl with the X-Ray Telescope (XRT; \citealt{Burrows2005}) on-board the \textit{Swift} satellite beginning at 19:32 UT on 30 August 2016.  Regular monitoring of the field in photon counting (PC) mode continued over the course of the next four months (PIs T. Holoien and B. Cenko).

No significant emission is detected in individual epochs (typical exposure times of $\approx$2\,ks). Using standard XRT analysis procedures (e.g., \citealt{Evans2009}), we place 90\% confidence upper limits ranging from $(3.8$--$12.1) \times 10^{-3}$ counts s$^{-1}$ in the 0.3--10.0\,keV bandpass over this time period.

Stacking all the XRT data obtained over this period together (58\,ks of total exposure time), we find evidence for a weak ($\approx 4\sigma$ significance) X-ray source at this location with a 0.3--10.0\,keV count rate of $(2.6 \pm 1.2) \times 10^{-4}$ counts s$^{-1}$.  

With only 15 source counts we have limited ability to discriminate between spectral models; however, with several photon energies detected above 1\,keV.  We derive response matrices for the stacked XRT observations using standard \textit{Swift} tools.  Adopting a power-law model for the spectrum with a photon index of 2 and accounting for line-of-sight absorption in the Milky Way \citep{Willingale2013MNRAS}, we find the measured count rate corresponds to an unabsorbed flux of 0.3--10.0\,keV flux of $4.6_{-2.0}^{+3.7} \times 10^{-15}$\,erg\,cm$^{-2}$\,s$^{-1}$.

At the distance of Mrk\,950, this corresponds to an X-ray luminosity of $L_{X} = 2.4_{-1.1}^{+1.9} \times 10^{39}$\,erg\,s$^{-1}$. Without additional information (e.g., variability and/or spectra), we cannot determine conclusively if this X-ray emission is associated with the transient iPTF16fnl, or if this is unrelated X-ray emission from the host nucleus (e.g., an underlying active galactic nucleus) or even a population of X-ray binaries or ultra-luminous X-ray sources.  However, the lack of evidence for ongoing star formation or AGN-like emission lines in the late-time optical spectra of Mrk\,950 (\S5) suggest an association with the tidal disruption event.

If this is indeed the case, the implied X-ray emission would be extremely faint, both in an absolute and a relative sense.  We can contrast, for example, with the X-ray emission observed from ASASSN-14li \citep{Holoien2016-14li,vanVelzen2016Sci}, with a peak luminosity approximately four orders of magnitude above that seen for iPTF16fnl.  Even sources with much fainter X-ray emission, such as ASASSN-15oi \citep{Holoien2016-15oi}, still outshine iPTF16fnl by more than a factor of 100.  



\subsection{Spectroscopic Observations} \label{sec:spectra}

\begin{figure*}[ht]
\includegraphics[width=\textwidth]{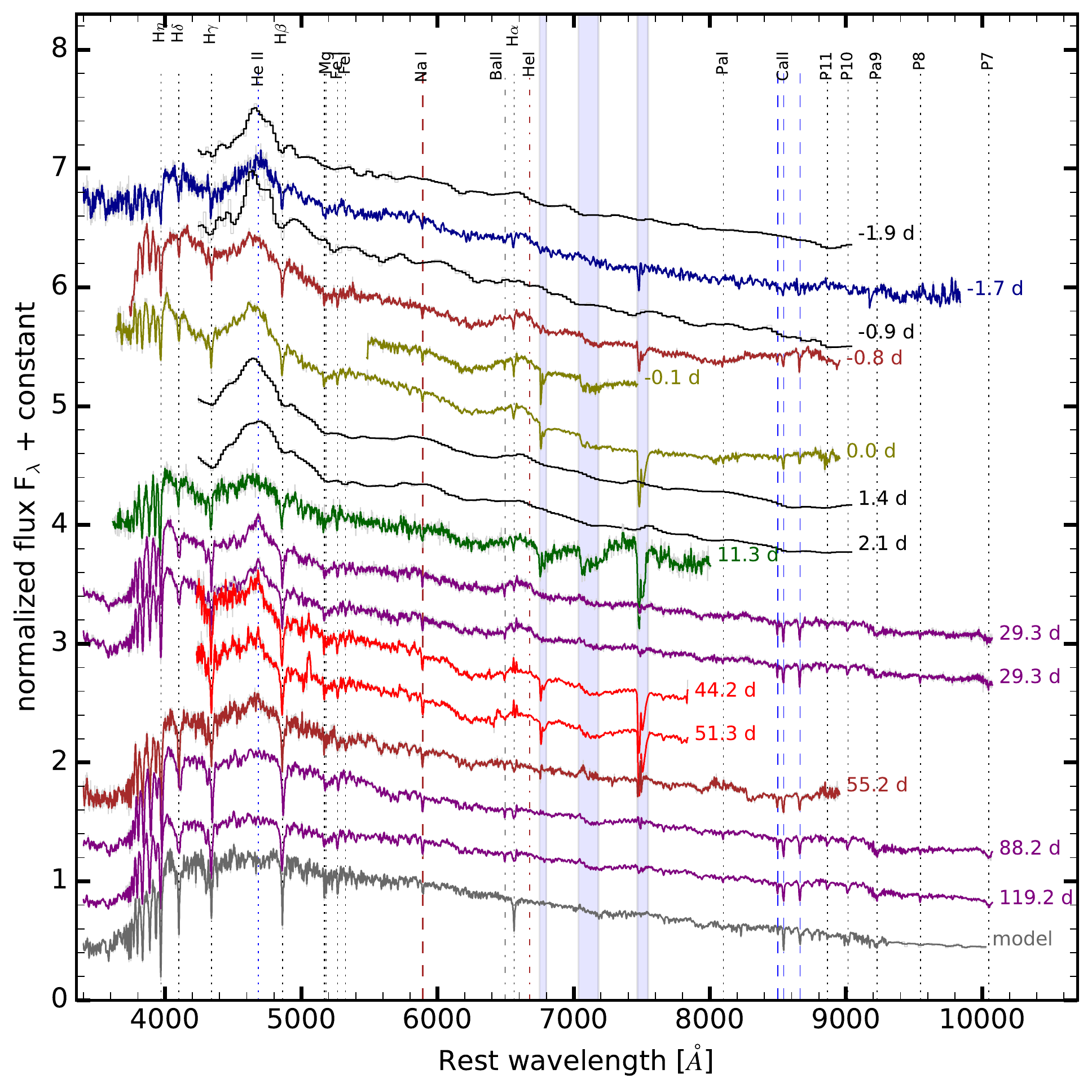} 
\caption{Spectral sequence of iPTF16fnl. The spectra are color coded by instrument: SEDM/P60 - black, DBSP/P200 - brown, 2\,m LOYDS/LCO - blue, LRIS/Keck - purple, Deveny/DCT - green, GMOS-N/Gemini North - red and GTC/OSIRIS-olive. Telluric bands are marked with blue shaded areas. The labels on the top of the panel correspond to the main identified lines both from the TDE and the host galaxy. The spectra have been binned using the average of 3 pixels and low S/N areas were excluded from the plot. The best-fit galaxy model is shown in the bottom with gray colour. }
\label{fig:spec}
\end{figure*}

Spectroscopic follow-up observations of iPTF16fnl have been carried out with numerous telescopes and instruments, summarized in the spectroscopic log in Table \ref{tab:speclog}. Figure \ref{fig:spec} shows the spectral sequence for iPTF16fnl, spanning three months. The spectroscopic data are made public via WISeREP \citep{YaronGal-Yam2012}.

Given the brightness of the galactic bulge relative to the TDE, the interpretation of the TDE spectrum poses some challenges. A noticeable feature is the strong host component in all the available spectra. Different slit widths, the variable seeing, different orientations of the slit during the acquisition of the data (generally taken at the parallactic angle) and the highly elongated geometry of the host galaxy, contribute to create a strong variation in the contribution from the host component, which appears to vary from one instrument to another.

Early epoch spectra ($<50$\,days), the most prominent emission lines correspond to broad \he2 and \halpha, shown in Figure \ref{fig:speclines}. These lines have an average FWHM of $\sim14,000$\,\kms and $\sim10,000$\,\kms respectively. The analysis and evolution of their profiles is further explored in Section \ref{sec:spec_analysis}.

Several narrow absorption lines, associated with the host galaxy, were identified. The region around \halpha contains an emission line that can be associated with [\io{N}{II}] at $\lambda$ 6583.
We also observed narrow absorption lines corresponding to \io{Ba}{II} at $\lambda\lambda$ 6496, the \io{Na}{I} D doublet at $\lambda\lambda$ 5889, 5896, \io{Mg}{I} $\lambda\lambda$ 5167, 5173, 5184, \io{Fe}{I} $\lambda\lambda$ 5266, 5324 \io{Ca}{II} is detected at $\lambda\lambda$ 3934, 3968 and as strong NIR triplet absorption at $\lambda\lambda$ 8498, 8542, 8662 \AA. 

\section{Analysis} \label{sec:analysis}

\subsection{SED and bolometric lightcurve}

\begin{figure}[ht!]
\includegraphics[width=0.5\textwidth]{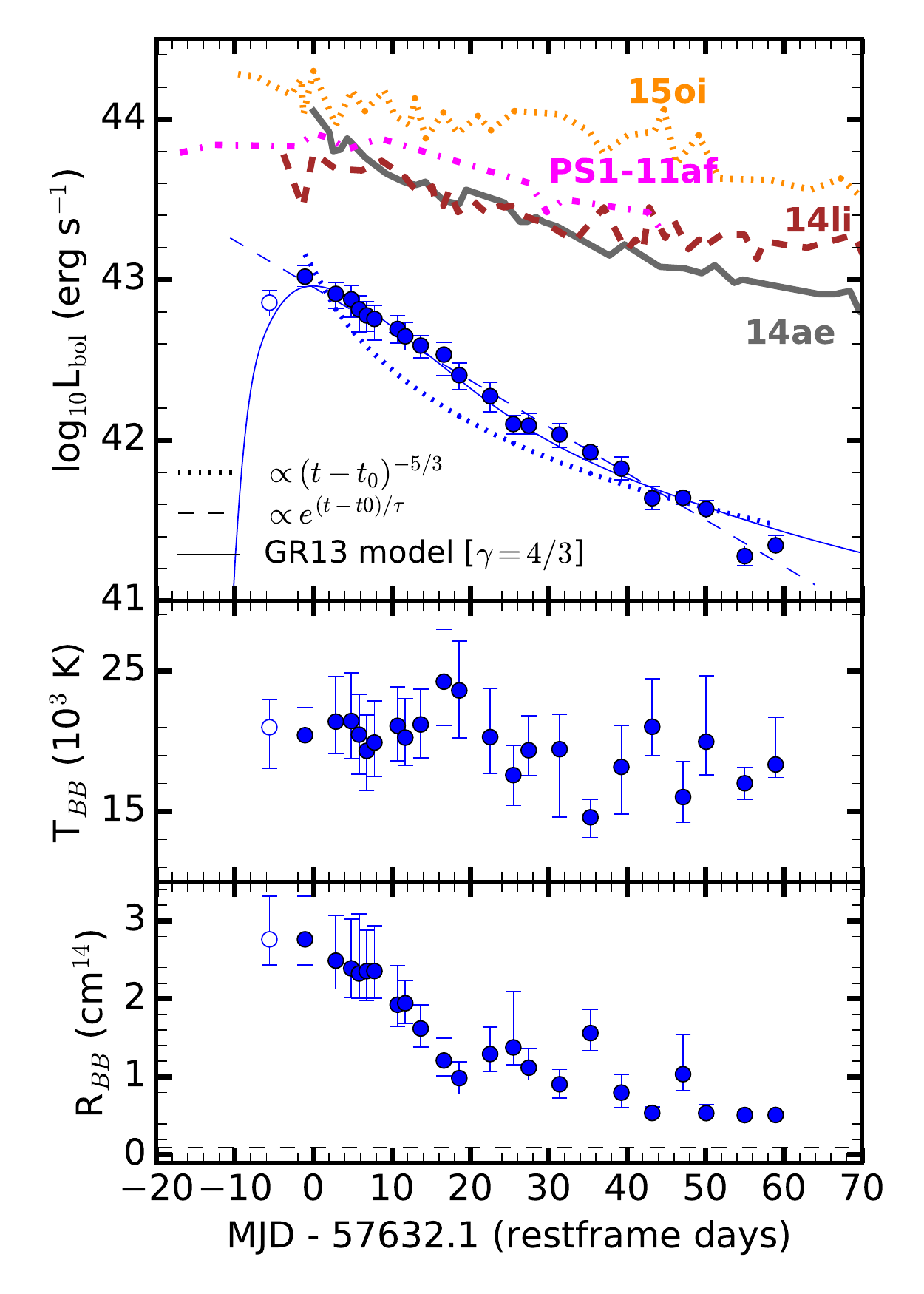} 
\caption{Top: Bolometric blackbody lightcurve for iPTF16fnl. Blue circles represent the fits with Galactic extinction correction only. The first (empty) data point was computed assuming an average blackbody temperature of 21,000\,K (average for the first 2 weeks) and scaling the flux to match the $g$-band magnitude. The dashed line shows the best fit to a power law of the form $L \propto e^{-(t-t_0)/\tau}$. The dotted line shows the best fit to a $L \propto (t-t_0)^{-5/3}$. A solid line shows the best fit to GR13 models. Thick lines represent a sample of fast-fading TDEs for comparison: PS1--11af: dot-dashed magenta \citep{Chornock2014ApJ}, ASASSN-14ae: solid gray \citep{Holoien2014-14ae}, ASASSN-14li: dashed brown \citep{Holoien2016-14li} and ASASSN-15oi: dotted orange \citep{Holoien2016-15oi}. The reference MJD for the objects is the discovery date or epoch of peak luminosity (whenever available). Middle: Temperature evolution. Bottom: Evolution of the blackbody radius. (A color version of this figure is available in the online journal.)}
\label{fig:bb}
\end{figure}

\textit{Swift} host subtracted $UVW1$, $UVM2$, $UVW2$ photometry and Palomar data were used to fit the object blackbody temperature and radius. We fit the fluxes derived from each band with spherical blackbody emission models using Markov Chain Monte Carlo (MCMC) simulations, based on the \texttt{Python} package \texttt{emcee} \citep{Foreman-Mackey2013PASP}. The blackbody bolometric luminosity, along with the best fit for the temperature and the radius are shown in Figure \ref{fig:bb}. Because only $g$-band measurements were available for our first detection epoch, we assumed that the luminosity follows a blackbody emission with a temperature of 21,000\,K (average for the first 2 weeks) and scaled the flux to match the $g$-band magnitude.

The bolometric luminosity at peak is $L_p~\simeq~(1.0\,\pm\,0.15) \times 10^{43}$\ergs. This is one order of magnitude lower than most of the optical TDEs (PS1-10jh, ASASSN-14ae, ASASSN-14li and ASASSN-15oi; see Figure \ref{fig:bb}). If compared to ASASSN-15lh, a TDE candidate from a rotating high-mass SMBH \citep{Leloudas2016NatAs} \footnote{This object is somehow controversial and has been initially interpreted as a superluminous supernova \citep{Dong2016Sci}}, the peak is two order of magnitude fainter.
Based on our $M_{BH}$ estimate, we derive its Eddington luminosity $L_{\rm{Edd}}=2.7^{+3.7}_{-1.6}\times 10^{44}$\ergs, implying that at peak, iPTF16fnl shines only 2$-$10\% of $L_{\rm{Edd}}$. Assuming a radiative efficiency of $\eta=0.1$, this translates into a peak accretion rate of $\dot{M}_{\rm{peak}}\sim 1.8\times10^{-3}$\,\Msun\,yr$^{-1}$ ($L=\eta \dot{M}c^2$). 

Integrating the available bolometric luminosity (see Figure \ref{fig:bb}), we find the radiated energy to be $E_R = (2.0\pm0.5) \times 10^{49}$\,erg. The accreted mass for this interval is $M_{acc} \sim 7.3\times10^{-5}$\,\Msun.

The rest-frame blackbody bolometric luminosity (Figure \ref{fig:bb}) is fit with the characteristic power law $L\propto ((t-t_0)/\tau)^{-5/3}$ and an empirically motivated exponential profile, $L \propto e^{-(t-t_0)/\tau}$, where $t_0$ and $\tau$ are free parameters. For the decaying part of the lightcurve, the exponential model fits the data better ($\chi^2$= 17 vs. 110), and best fit parameters are $\tau_0\simeq15$ and $t_0\simeq-6$. For comparison, the decay for other optical TDEs is slower: ASASSN-14ae had $\tau$ = 30\,days, whereas  ASASSN-15oi and ASASSN-14li faded on timescales of 46.5 and 60 days respectively. Continued, high quality photometric monitoring would be required to draw conclusive results on long-term evolution, beyond the initial fading stage. 

We estimate a lower limit for the time from disruption to peak light $t_{\rm{peak}} \geq 11$\, days from the bolometric lightcurve. We select the measurements at $\pm$10\,day from the peak in $g$-band and fit the luminosity with a 2 degree polynomial. For the raising part of the lightcurve, we use our only available $g$-band measurement. While this approach is widely used in estimating the explosion time for supernovae, the emission mechanism for TDEs is different and therefore it only yields to lower limits, as seen when applied to PS1-10jh.  

Assuming that our bolometric lightcurve traces the rate of mass falling into the black hole, $\dot{M}(t)$, we use the lightcurve models from \cite{Guillochon2013ApJ} (hereafter GR13), scaling them to the peak accretion mass rate and time to peak. The lightcurves are defined for a range of impact parameters $0.5 \leq \beta \leq 2.5$, where $\beta$ is defined as the depth of the encounter $\beta =R_T/R_p$, $R_T$ is the tidal disruption radius and $R_p$ the pericentre radius. We impose a $M_{BH}=2\times 10^6$\, \Msun, but we leave the mass and radius of the disrupted star as free parameters. Our best fit corresponds to a star with polytropic index $\gamma=4/3$ (fully radiative) with $M_{*}\sim 0.03$\,\Msun, $R_{*} \sim 0.3$\,\Rsun and a depth of the encounter comparable to the disruption radius ($\beta \simeq 1$). The lightcurve for the best fit model is shown in Figure \ref{fig:bb}. We obtain $t_{peak} \sim 11$\, days, comparable with our previous naive estimate. If we impose that low mass stars are fully convective and fit for an object with $\gamma=5/3$, we find a relatively good fit for a partial disruption ($\beta \sim 0.6$) and a similar value of $M_{*}\sim 0.06$\,\Msun, although in this case $t_{peak}$ is shorter than our observations suggest. Although these values illustrate that the disrupted object was likely a low-mass star, detailed modeling of the event would be required to draw quantitative results.

The blackbody model has an average temperature of T$_{BB}=19,000\pm$2000\,K, which  does not vary significantly over time, as shown Figure \ref{fig:bb}. At later epochs, the increased uncertainties in the host subtraction lead to increased scatter in T$_{BB}$. Given the uncertainty of the extinction in the host, these values can be assumed as lower limits. The model blackbody radius, R$_{BB}$ starts at $\sim$2.5$\times10^{14}$\,cm, linearly declines for the first twenty days and then flattens to $5\times10^{13}$\,cm. These radii are much larger than the Schwarzschild radius $r_{Sch} \sim 6\times10^{11}$\,cm of the nuclear black hole. In comparison, we note that the tidal disruption radius of such SMBH for a main sequence Solar-like star is $R_T\simeq1\times10^{13}$\,cm. Photospheric emission at radii larger than $R_T$ is commonly observed for the optical sample of TDEs. This has been attributed to the existence of a reprocessing layer at larger radii, which re-emits the X-ray and UV in optical bands \citep{LoebUlmer1997ApJ,Guillochon2014ApJ}. Alternatively, the emission mechanism may originate from the energy liberated by shocks between streams in the apocenter, during the formation of the accretion disk \citep{Piran2015ApJ}. Such an optically thick layer, mainly formed of stellar debris, is associated with the origin of the emission line signature for optical TDEs \citep{Roth2016ApJ,MetzgerStone2016MNRAS}.

\subsection{Spectroscopic analysis}\label{sec:spec_analysis}

\begin{figure}[ht]
\includegraphics[width=0.5\textwidth]{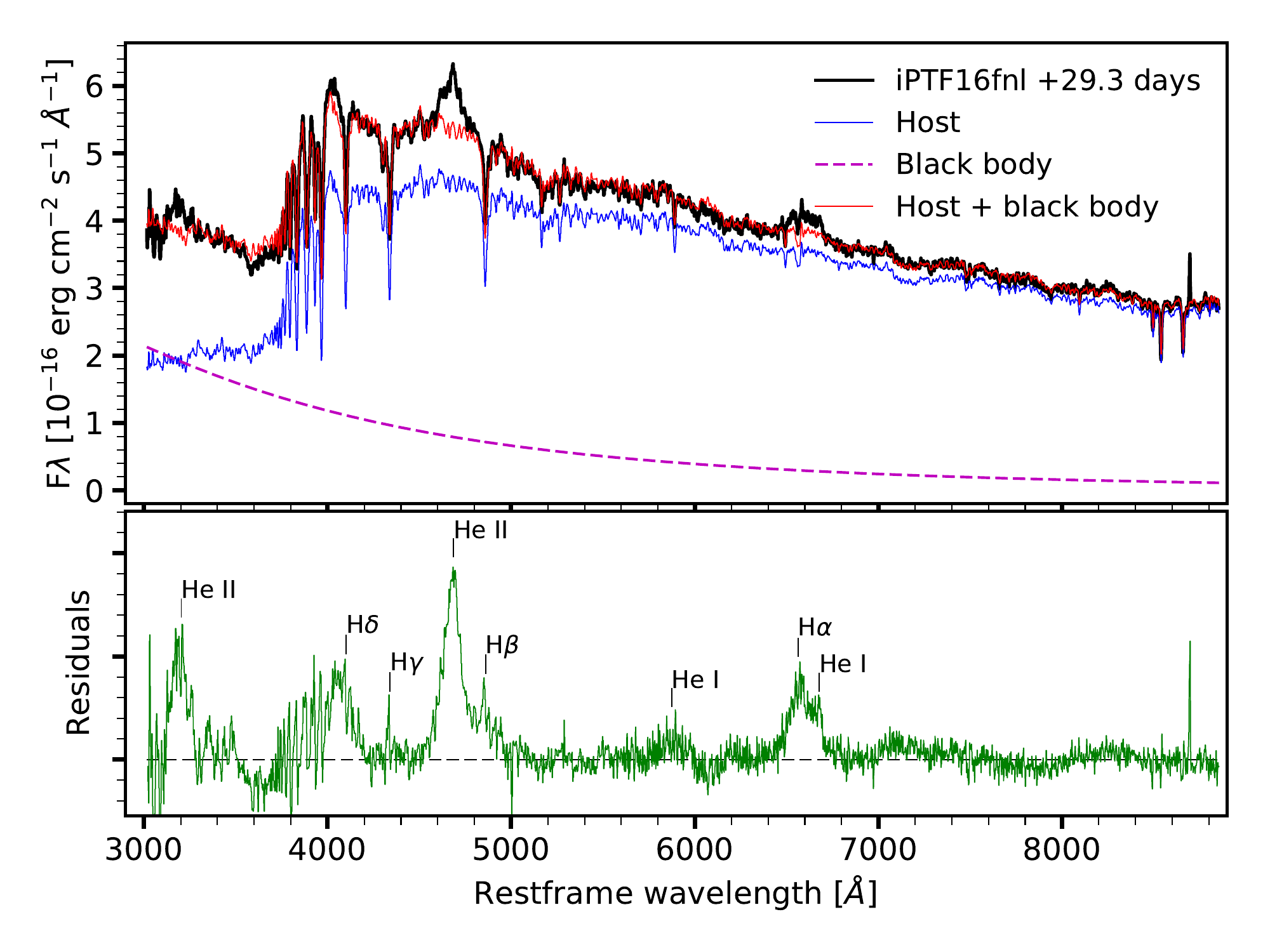} 
\caption{Example of host subtraction. The original spectrum, taken a +29.3 days (black thick line) was fit with a combination of host spectrum (blue solid line) and a blackbody fit (magenta dashed line). The best fit is shown with a red line. The residuals (corrected with a 2 degree polynomial) are shown in the lower panel. We mark the relevant emission lines. \io{He}{II} lines are clearly identified at $\lambda\lambda$ 3203 and 4686. \io{He}{I} lines are present at $\lambda\lambda$  5875 and 6678. (A color version of this figure is available in the online journal.)}
\label{fig:residuals}
\end{figure}

\begin{figure*}[ht]
\includegraphics[width=\textwidth]{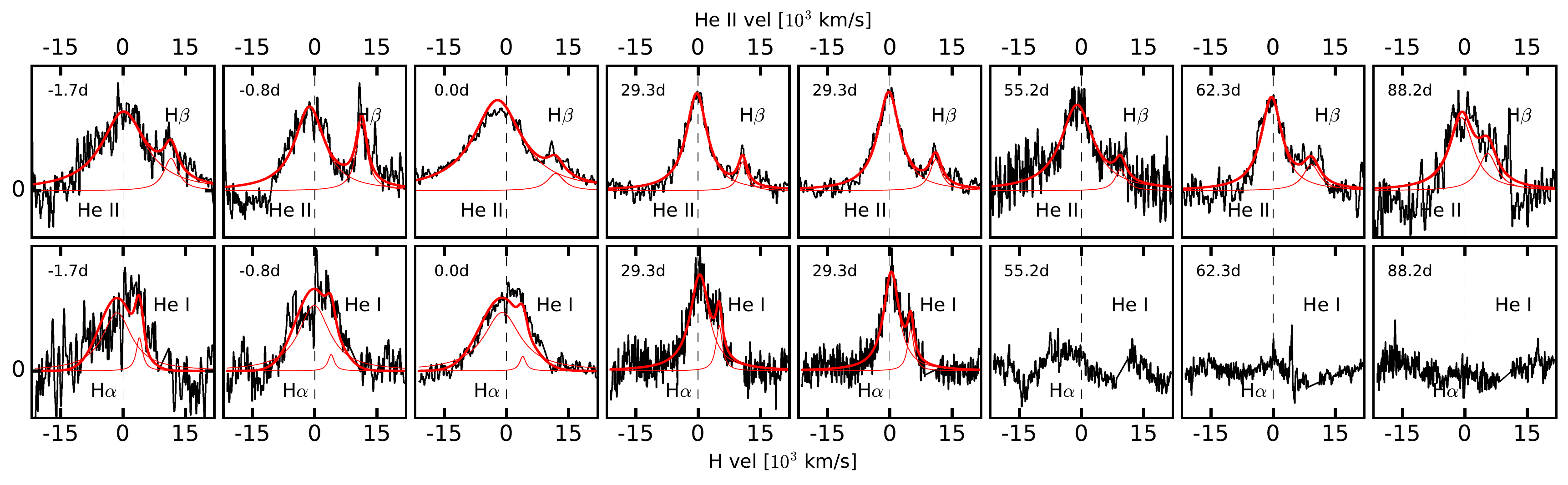} 
\caption{Residual normalized spectrum showing the line region around \he2 $\lambda$4686 (top row) and \halpha (bottom row) lines for the higher S/N spectra of iPTF16fnl. In addition, the location of H$\beta$ (top) and He I (bottom) lines are shown. Telluric regions, shown with shaded areas, were excluded from the fit. The first three epochs of H$\alpha$ were fit using a Gaussian model, as the Lorentzian provided a worse fit. All the other lines were fit using a linear background and a Lorentzian line profile. We could not find a good fit for the last three epochs of \halpha, but the spectrum is included for completeness. (A color version of this figure is available in the online journal.)}
\label{fig:speclines}
\end{figure*}

The early time spectrum of iPTF16fnl is dominated by blue continuum radiation and the characteristic broad \he2 $\lambda$4686 line. The emission around 6500\,\angstrom can be attributed to \halpha, although the \io{He}{I} $\lambda$6678 line is also detected. In the region around He II, we also detect H$\beta$ in emission. However the strong host contribution makes its identification challenging at late times.
In our analysis, we use the late time host spectrum (+119.2 days) as our template. We select the highest signal-to-noise (S/N) spectra, and fit them with a combination of host and a blackbody continuum, as show in Figure \ref{fig:residuals}. On the residual spectrum, we fit a line model using the \texttt{python} package \texttt{lmfit} (Non-Linear Least-Squares Minimization and Curve-Fitting for Python). After masking the regions affected by telluric absorption, \he2 + H$\beta$ and \halpha + \io{He}{I} lines are fit using two component Lorentzian model, in order to derive the width (FWHM) and central location of the emission. The results, plotted as insets, are shown in Figure \ref{fig:speclines}. The fluxes for each line are derived from the best fit model and shown in Table \ref{tab:lineflux}. As discussed in \cite{Brown2017arXiv}, if all the flux in the \io{He}{II} line would be attributed to recombination produced by black body photoionizing radiation, the observed flux of $>10^{40}$\,erg\,s$^{-1}$ would require black body radiation with temperature $\sim4\times10^4$\,K, which is higher than our fit, requiring an additional energy source to power this line.

Around peak, the \he2 lines appear to have higher velocity, showing an average value of FWHM$_{\rm{HeII}}\simeq 14,000\,\pm 3,000$\,\kms, in contrast to the \halpha line, with FWHM$_{H\alpha}\simeq 10,000 \pm 500$\,\kms. At +30 and +45 days after peak, the FWHM narrows down to $8,500\pm 1500$\,\kms for \he2 and $6,000\pm 600$\,\kms for \halpha.
The center of the lines appears constant within the scatter for the first 90 days: for \he2, the lines appear marginally blueshifted with velocity of $-700\pm 700$\,\kms, while the \halpha lines appear to be consistent with the reference wavelength, with a shift in velocity of $-800\pm 1200$\,\kms. 

\begin{table}
\begin{minipage}{1.\linewidth}
\centering

\begin{small}

\caption{Flux values for H$\alpha$ and \io{He}{II} 4868\AA for the lines shown in Figure \ref{fig:speclines}. The values were derived from the best model fit line profile. From the fit uncertainties, we estimate errors of 40\% and 30\% of the total flux for H$\alpha$ and \io{He}{II} lines respectively.}
\begin{tabular}{cccc}
\hline
MJD & Phase  & H$\alpha$ & \io{He}{II} \\ 
(d) & (d) & (erg \,s$^{-1}$) & (erg \,s$^{-1}$) \\ \hline
57630.4 & $-$1.7 & 2.1$\times10^{40}$ & 8.9$\times10^{40}$ \\
57631.0 & $-$0.8& 3.8$\times10^{39}$ & 1.5$\times10^{40}$ \\
57631.3 & 0.0 & 6.6$\times10^{39}$ & 8.3$\times10^{39}$ \\
57661.4 & 29.3 & 3.0$\times10^{39}$ & 3.0$\times10^{39}$\\
57661.4 & 29.3 &2.1$\times10^{39}$ & 2.5$\times10^{39}$\\
57687.3 & 55.2& -- & 1.1$\times10^{40}$ \\
57694.4 & 62.3& -- & 2.7$\times10^{39}$ \\
57720.3 & 88.2& -- & 1.1$\times10^{39}$ \\
\hline
\end{tabular}
\label{tab:lineflux}
\end{small}
\end{minipage}
\end{table}

\begin{figure}[h]
\includegraphics[width=\columnwidth]{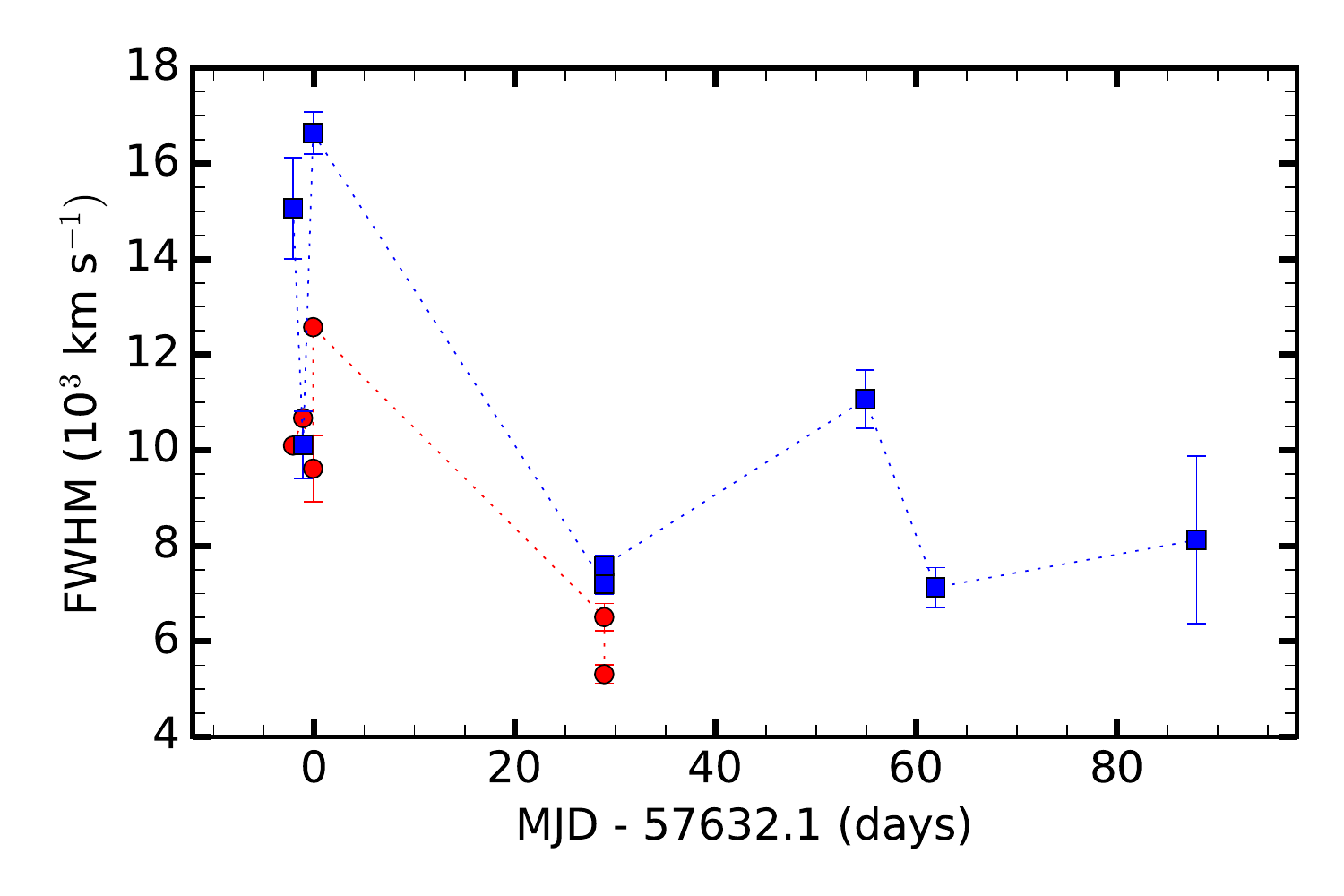} 
\caption{Top: Evolution of FWHM for \io{He}{II} $\lambda\lambda$4686 \AA~ ~line (blue squares) and \halpha (red circles) vs. phase of the spectrum. The last three epochs of \halpha do not have a reliable measurement, and therefore are excluded form the figure. (A color version of this figure is available in the online journal.)}
\label{fig:lines}
\end{figure}

\begin{figure}
\hspace{-0.6cm}
\includegraphics[width=1.1\columnwidth]{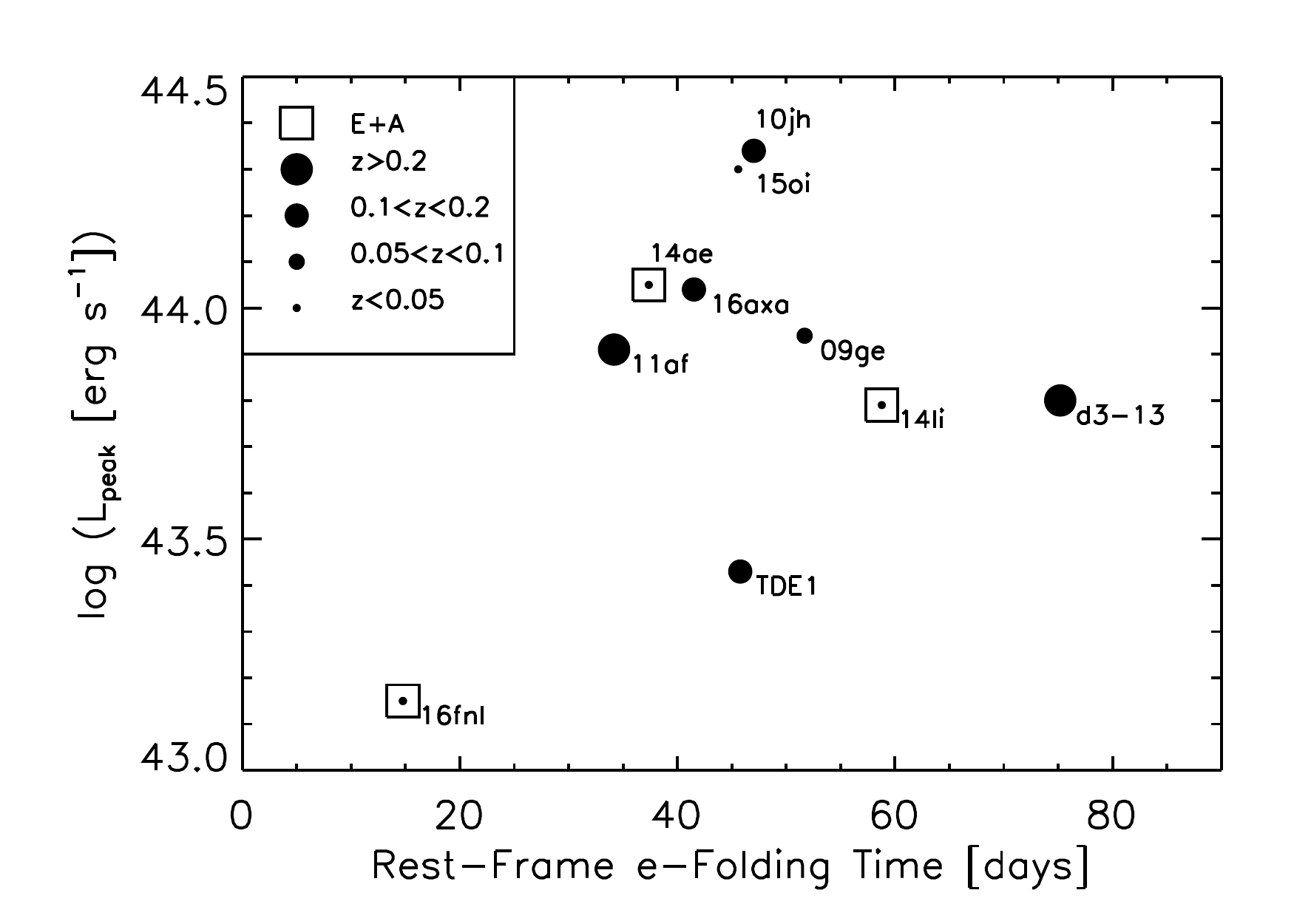} 
\caption{Comparison of the peak luminosity and decay time of iPTF16fnl with a sample of optical TDE from literature. The dot size encodes the redshift of the host galaxy. An external circle symbolizes the classification of the host galaxy as a post-starburst E+A galaxy. The optical TDE sample is based on published data: \cite{Gezari2008ApJ,vanVelzen2011ApJ,Gezari2012Natur,Chornock2014ApJ,Arcavi2014,Holoien2016-14li,Holoien2014-14ae,Holoien2016-15oi} and \cite{Hung2017arXiv}.}
\label{fig:comparison}
\end{figure}

\section{Discussion}\label{sec:discussion}

iPTF16fnl is the faintest and fastest event in the current sample of optically discovered TDEs. Assuming our extinction estimation method is accurate, its luminosity at peak is one order of magnitude lower than any other optical/UV TDE discovered so far. Its timescale, as shown in Figure \ref{fig:comparison}, also makes it as an outlier among the existing sample. 

The host of iPTF16fnl is another example of a TDE in a post-starburst galaxy, further linking the propensity of TDEs to such galaxies \citep{Arcavi2014,French2016ApJ}. Moreover, E+A galaxy hosts seem to be exclusive for the lowest redshift TDEs ($z < 0.05$) (see Figure \ref{fig:comparison}). The origin of the burst could be associated with a merger episode, as discussed in the case of ASASSN-14li \citep{Prieto2016}. The violent relaxation in the stellar orbits could enhance the rate of captures, as stars can undergo encounters that will scatter them towards the SMBH. 

\begin{figure}
\hspace{-0.6cm}
\includegraphics[width=1.1\columnwidth]{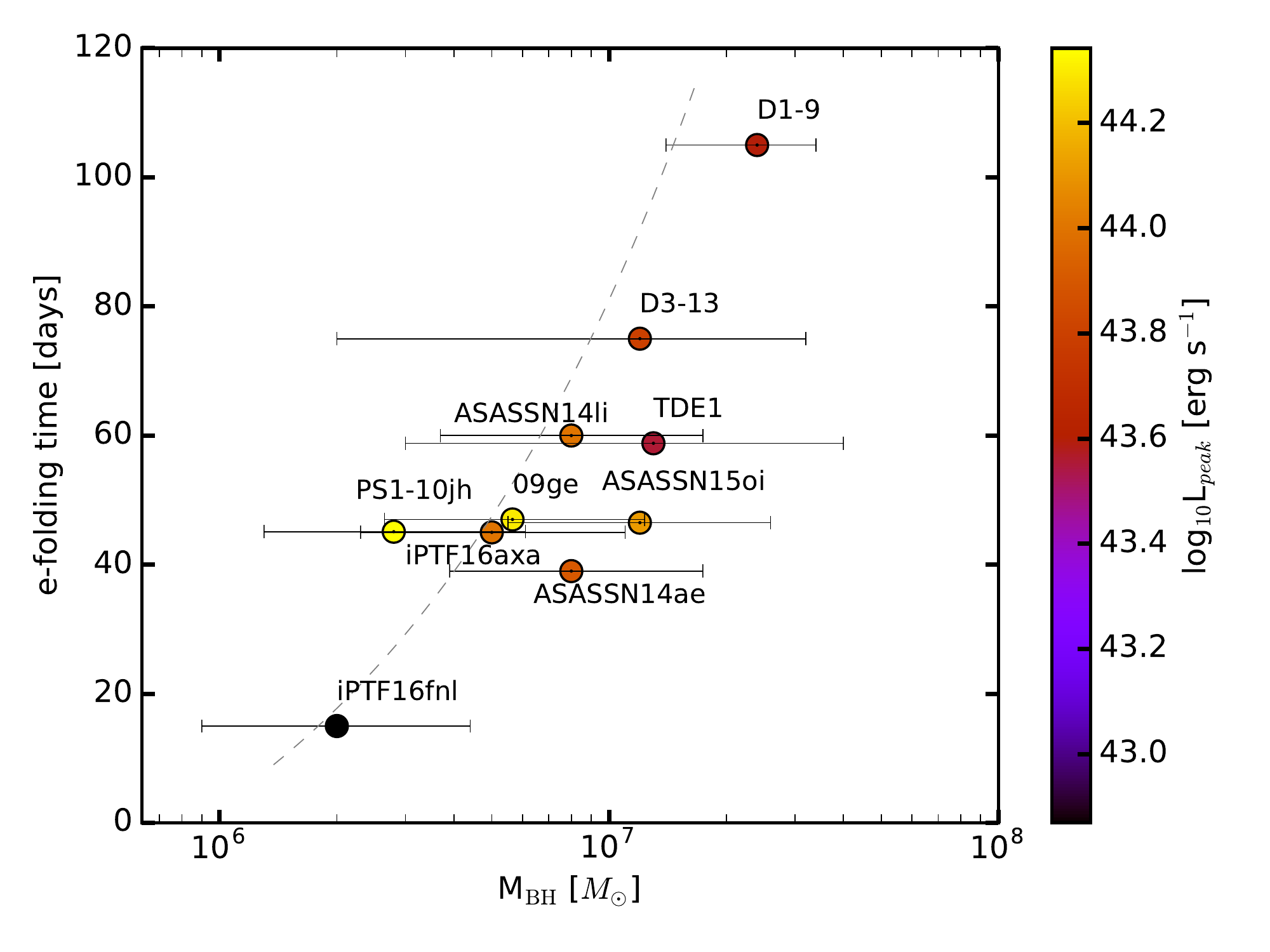} 
\caption{Mass of the host galaxy SMBH compared to the e-folding timescale for a sample of optical TDE. The dot color encodes the TDE peak luminosity. The peak luminosities were derived from literature: D1-9, D3-13 \cite{Gezari2008ApJ,Gezari2009ApJ}, PS1 \citep{Gezari2012Natur}, PS1-11af \citep{Chornock2014ApJ}, ASASSN-14ae, ASASSN-14li, ASASSN-15oi \citep{Holoien2016-14li,Holoien2014-14ae,Holoien2016-15oi}. The bolometric luminosities for TDE1 \cite{vanVelzen2011ApJ} and 09ge \citep{Arcavi2014} were derived by scaling the reported blackbody temperature emission to match the reported $M_g$. We assumed the standard dispersion in the \cite{McConnell2013ApJ} relation whenever uncertainties for $M_{BH}$ were not reported.  A tentative correlation $t \propto (M_{BH})^{1/2}$ is provided to guide the eye \citep{Guillochon2013ApJ}. (A color version of this figure is available in the online journal.) }
\label{fig:compare_timescales_mbh}
\end{figure}

Lower SMBH masses ($<10^7$\Msun) can increase the number of deeper encounters \citep{Kochanek2016MNRAS}, allowing for disruptions with smaller pericenter radius, $R_p$. However, theoretical works \citep{Guillochon2013ApJ,StoneSariLoeb2013MNRAS} only show a weak correlation between the impact parameter $\beta$, and the peak of the flare. Therefore, the low luminosity and fast timescales shall be attributed to a lower mass black hole and/or lower mass for the disrupted star. Using our fit to TDE lightcurves from \cite{Guillochon2013ApJ} with our estimated $M_{BH} \sim 2 \times 10^6$\Msun, we obtain the mass of the disrupted star to be $M_{*} \sim 0.03$ for $t_{peak}$ value of 11\,days.

iPTF16fnl has clearly faster decay timescales than other TDEs, but also lower $M_{BH}$. Figure \ref{fig:compare_timescales_mbh} shows a comparison of the e-folding timescale for iPTF16fnl and other optical TDEs, computed from exponential decay models, and the galaxy $M_{BH}$. There seems to be a trend between these two values for the optical/UV TDE population, in general agreement with the theoretical scaling between fallback timescale and black hole mass, $t \propto M_BH^{1/2}$. As a cautionary note, while literature generally reports $M_{BH}$ based on bulge mass/luminosity, our best measurement is based on the $M-\sigma$ relation, although the bulge luminosity method yielded to similar results. Figure \ref{fig:comparison}  shows that the most luminous flares ($L_{\rm{bol}}>10^{44}$ \ergs) tend to fade on intermediate timescales, $\sim$50 days. However, there does not seem to be an evident correlation between the peak luminosity and the black hole mass, as discussed in \cite{Hung2017arXiv} (see their figure 15).

The tension between theoretical prediction of TDE rates and the ones inferred from observations is an active field of research. While it is difficult to explain the differences in terms of host galaxy properties  \citep{StoneMetzger2016MNRAS}, an observational bias towards the brighter events seems to offer a more plausible explanation. The discovery of iPTF16fnl has consequences for previous optical searches for nuclear tidal disruptions. In fact, its peak absolute magnitude $M_g=-17.2$\,mag, and fast decay timescales, may mimic the behavior of a SN exploding close to the galaxy nucleus. Therefore, such faint events may have gone unnoticed in searches for bright ($M_g \sim -20$ mag) nuclear flares \citep{Arcavi2014}.

Systematic searches using the colour (including UV) and location of the transient, rather than its absolute magnitude, will increase our sensitivity to fainter flares. Consistent candidate selection using future surveys such as ZTF or LSST will allow us to explore the full luminosity function of tidal disruption flares. Spectroscopic confirmation of the candidates will be essential to identify this faint population. Dedicated instruments for transient classification such as SEDM will become the big players in this new era.

\newpage
\section{Conclusions}\label{sec:conclusions}

We have presented the discovery and follow-up data for iPTF16fnl, a TDE candidate discovered by the iPTF survey on 2016 August 29th. The real-time image-subtraction pipeline and rapid spectroscopic classification allowed us to initiate a timely follow-up campaign. The photometric and spectroscopic signatures of iPTF16fnl are consistent with the sample of previous optically selected TDEs. As observed in other TDEs, the object shows very strong emission in UV wavelengths, with a $T_{BB} \simeq 19,000$\,K. The temperature does not show strong evolution and the decrease in luminosity is best explained as a decrease in the size of the radiating region. In agreement with previous work, the size of this region, defined by its photospheric radius, is also about an order of magnitude larger than $R_T$. The early times, the spectroscopic signature of iPTF16fnl is dominated by \he2 and hydrogen lines, although we also detect emission from He I. After two months after peak light, most of the lines have faded. The exception is \he2, which can be identified with a relatively constant FWHM of $\sim$7,000\,\kms.

iPTF16fnl is remarkable in three ways: it is the nearest well studied optical/UV TDE (66.6\,Mpc), and it has one of the shortest exponential decay timescales (about 15 days) and one of the lowest peak luminosities, $L_p~\simeq~(1.0\,\pm\,0.15) \times 10^{43}$\ergs. Also, its host galaxy has the lowest $M_{BH}$ among the optical sample of TDEs. Although this could justify the fast decay, its low luminosity may be related to the disruption of a lower mass star, or even a partial disruption. 

Our work demonstrates that TDEs cover a wide range of luminosities and timescales. Current and future surveys, such as ATLAS, PS-1, ZTF and LSST, will provide large numbers of events with exquisite temporal coverage. Multifrequency follow-up of the candidates will lead to a better understanding of the underlying emission mechanism, hopefully leading to a unified multifrequency emission model. Large samples obtained with well understood selection criteria will be key to the study of TDE demographics, providing a unique link between theoretical studies of SMBH astrophysics and observations.

\scriptsize
\bibliographystyle{aasjournal}
\bibliography{main}

\facility{DCT, Gemini North, GTC, Hale, JMCT, Keck:I, Keck:II, LCOGT, PO:1.2\,m, PO:1.5\,m, PS1, VLT  }

\software{AstroPy, EMCEE \citep{Foreman-Mackey2013PASP}, IRAF, LMFIT, pPXF \citep{2016arXiv160708538C}, PTFIDE \citep{Masci2017PASP}, PYRAF, UVOTSOURCE}.

\section*{Acknowledgments} \label{sec:acknowledgements}

\tiny
We thank M. Urry for the use of her programme time to obtain target of opportunity observations of iPTF16fnl with DBSP on 200-inch telescope on Palomar. We also thank Nathaniel Roth, Sjoert Van Velzen and Nicholas Stone for useful comments and discussions, and James Guillochon for his support with \texttt{TDEFit}.

This work was supported by the GROWTH project funded by the National Science Foundation under Grant No 1545949. We thank the Gemini Fast Turnaround program (PI: T. Hung). S.G. is supported in party by NSF CAREER grant 1454816 and NASA Swift Cycle 12 grant NNX16AN85G.
A.Y.Q.H. was supported by a National Science Foundation Graduate Research Fellowship under Grant No. DGE‐1144469. 
A.J.C.T. acknowledges support from the Spanish Ministry Project
AYA 2015-71718-R.
Support for I.A. was provided by NASA through the Einstein Fellowship Program, grant PF6-170148.
G.H. is supported by the National Science Foundation (NSF) under Grant No.~1313484. This work makes use of observations from the LCO network.
G.H. is supported by the National Science Foundation (NSF) under Grant No.~1313484. This work makes use of observations from the LCO network.

We acknowledge the use of public data from the Swift data archive.
The CSS survey is funded by the National Aeronautics and Space
Administration under Grant No. NNG05GF22G issued through the Science Mission Directorate Near-Earth Objects Observations Program.  The CRTS survey is supported by the U.S.~National Science Foundation under grants AST-0909182.
LANL participation in iPTF was funded by the US Department of Energy as part of the Laboratory Directed Research and Development program. Part of this research was carried out at the Jet Propulsion Laboratory, California Institute of Technology, under a contract with the National Aeronautics and Space Administration. 
Part of this work is based on observations obtained at the Gemini observatory under the Fast Turnaround program.
The Gran Telescopio Canarias (GTC) operated on the island of La Palma at the Spanish Observatorio del Roque de los Muchachos of the Instituto de Astrofisica de Canarias.
This publication makes use of data products from the Wide-field Infrared Survey Explorer, which is a joint project of the University of California, Los Angeles, and the Jet Propulsion Laboratory/California Institute of Technology, funded by the National Aeronautics and Space Administration.
The James Clerk Maxwell Telescope is operated by the East Asian Observatory on behalf of The National Astronomical Observatory of Japan, Academia Sinica Institute of Astronomy and Astrophysics, the Korea Astronomy and Space Science Institute, the National Astronomical Observatories of China and the Chinese Academy of Sciences (Grant No. XDB09000000), with additional funding support from the Science and Technology Facilities Council of the United Kingdom and participating universities in the United Kingdom and Canada.
VAF is supported by the Russian Science Foundation under grant 14-50-00043.
\newpage
\appendix \label{appendix}

\renewcommand{\tabcolsep}{0.1cm}

\begin{table*}
\begin{minipage}{1.\linewidth}
\begin{small}
\caption{Log of spectroscopic observations of iPTF16fnl.}
\centering
\begin{tabular}{cccccccc}
\hline
MJD & Slit$^a$ & Telescope+Instrument & Grism$/$Grating & Dispersion & Resolution$^b$ & Exposure   \\ 
   (d) &  (arcsec)  			&  					   &	& (\angstrom/pix)		& \angstrom km/s) &	(s)	     \\ \hline
 \hline
57630.2 & 3.0 & P60+SEDM & none & 25.5 & 540 (2900\,\kms) & 900 \\ 
57630.4 & 2.0 & LCO 2-m+FLOYDS & -- & 1.7 & 15.3 (820\,\kms) & 2700 \\ 
57631.0 & 1.0 & GTC+Osiris & 1000B+2500I & 2.12+1.36 & 7.1 (380\,\kms) & 300 \\ 
57631.0 & 1.0 & GTC+Osiris & 2500R & 1.04 & 7.1(380\,\kms) & 300 \\ 
57631.2 & 2.2 & P60+SEDM & none & 25.5 & 504 (2700\,\kms) & 900 \\ 
57631.3 & 1.5 & P200+DBSP & 600/4000 & 1.5 & 10.5 (560\,\kms) & 600 \\ 
57633.5 & 3.6 & P60+SEDM & lenslet arr. & 25.5 & 672 (3600\,\kms)& 1800 \\ 
57634.2 & 3.0 & P60+SEDM & lenslet arr. & 25.5 & 613 (3300\,\kms)& 1200 \\ 
57636.2 & 3.0 & P60+SEDM & lenslet arr. & 25.5 & 659 (3500\,\kms)& 1200 \\ 
57637.2 & 3.6 & P60+SEDM & lenslet arr. & 25.5 & 589 (3200\,\kms)& 1200 \\ 
57638.2 & 3.0 & P60+SEDM & lenslet arr. & 25.5 & 556 (3000\,\kms)& 1200 \\ 
57643.22 & 1.0 & VLT Kueyen+UVES & 437+860 & 0.030+0.050 & 0.15 (8\,\kms) & 2x1500 \\
57643.25 & 1.0 & VLT Kueyen+UVES & 346+580 & 0.024+0.034 & 0.15 (8\,\kms) & 2x1500 \\
57643.4 & 1.5 & DCT+Deveny & 300 & 2.17 & 6.9 (370\,\kms) & 600 \\ 
57661.4 & 1.0 & Keck I+LRIS	& 400/3400+400/8500 & 2.0 & 6.5 (350\,\kms) & 600 \\
57661.4 & 1.0 & Keck I+LRIS	& 400/3400+400/8500 & 2.0 & 6.5 (350\,\kms) & 600 \\
57676.3 & 1.0 & GeminiN+GMOS & -- & 1.35 & 8.1 (435\,\kms) & 600 \\ 
57683.4 &  1.0 & GeminiN+GMOS & -- & 1.35 & 8.1 (435\,\kms) & 1200 \\ 
57687.3 & 1.5 & P200+DBSP & 600/4000 & 1.5 & 10.5 (560\,\kms) & 600 \\ 
57694.4 & 1.0 & Keck I+LRIS	& 400/3400+400/8500 & 2.0 & 7.2 (390\,\kms) & 1350 \\
57720.3 & 1.0 & Keck I+LRIS	& 400/3400+400/8500 & 2.0 & 6.5 (350\,\kms) & 1800 \\
57751.3 & 1.0 & Keck I+LRIS	& 400/3400+400/8500 & 2.0 & 6.5 (350\,\kms) & 1800 \\
\hline
\end{tabular}
\begin{tablenotes}
    \item[\textdagger]$^a$ For the IFU, the extraction radius (in arcsec) is indicated.
    \item[\textdagger]$^a$ Measured using the FWHM of $\lambda$ 5577 \io{O}{I} sky line.
    \end{tablenotes}
\label{tab:speclog}
\end{small}
\end{minipage}
\end{table*}

\clearpage

\renewcommand{\tabcolsep}{0.05cm}
\begin{deluxetable*}{rccccccccccc} 
\tabletypesize{\scriptsize} 
\tablewidth{0pt} 
\tablecaption{Optical and UV photometry of iPTF16fnl in AB magnitude system. \\
These are the originally measured magnitudes with \textit{Swift} and difference imaging piplines. For P48+CFH12k, the $r$-band column contains measurements in Mould-$R$ filter system.\\
These measurements are not corrected for Galactic extinction. Table \ref{table:phot} is published in its entirety in the machine-readable format. A portion is shown here for guidance regarding its form and content. \label{table:phot}} 
\tablehead{ 
   \colhead{MJD}     &  \colhead{Telescope}    &  \colhead{$UVW1$} &  \colhead{$UVM2$} &  \colhead{$UVW2$} &  \colhead{$U$} &  \colhead{$B$} &  \colhead{$V$} &  \colhead{$u$} &  \colhead{$g$} &  \colhead{$r$} &  \colhead{$i$} \\ 
\colhead{ (days) } & \colhead{ + Instrument} &  \colhead{(mag)}&  \colhead{(mag)}&  \colhead{(mag)}&  \colhead{(mag)}&  \colhead{(mag)}&  \colhead{(mag)}&  \colhead{(mag)}&  \colhead{(mag)}&  \colhead{(mag)}&  \colhead{(mag)}
} 
\startdata 
 57626.4 & P48+CFH12k & -- & -- & -- & -- & -- & -- & -- & 17.50$\pm$0.07 & -- & --\\
 57629.4 & P48+CFH12k & -- & -- & -- & -- & -- & -- & -- & 17.23$\pm$0.09 & 17.54$\pm$0.08 & --\\
 57630.8 & Swift+UVOT & 16.83$\pm$0.04 & 16.92$\pm$0.04 & 16.61$\pm$0.04 & 16.35$\pm$0.04 & 15.68$\pm$0.03 & 15.19$\pm$0.04 & -- & -- & -- & --\\
 57631.7 & P60+SEDM& -- & -- & -- & -- & -- & -- & 17.13$\pm$0.09 & 17.17$\pm$0.07 & 17.42$\pm$0.12 & --\\
... & ... & ... & ... & ... & ... & ... & ... & ... & ... & ... & ...
 \enddata 
\end{deluxetable*}

\end{document}